\documentclass[twocolumn,showpacs,showkeys,pre,floatfix]{revtex4}

\usepackage{bm}
\usepackage[dvips]{graphicx}
\usepackage{color}
\usepackage[usenames,dvipsnames]{xcolor}
\usepackage{amsmath, amsthm, amssymb}
\usepackage{theorem}
\usepackage{ascmac}
\usepackage{xcolor}
\usepackage{mathrsfs}
\usepackage{xspace}

\newcommand{\singlefiguresize}{0.85\columnwidth}
\newcommand{\doublefiguresize}{1.85\columnwidth}
\newcommand{\VEC}[1]{\overrightarrow{#1}}
\newcommand{\cav}[1]{\left\langle #1 \right\rangle_{\beta}}
\newcommand{\eav}[1]{\left\langle\!\!\!\:\left\langle #1 \right\rangle\!\!\!\:\right\rangle}
\newcommand{\quasi}{{\mathrm{QE}}}

\newcommand{\qav}[1]{\left\langle\!\!\!\:\left\langle #1 \right\rangle\!\!\!\:\right\rangle_{\quasi}}
\newcommand{\tav}[1]{\overline{#1^\quasi}}
\newcommand{\QE}{QE\xspace}
\newcommand{\fe}{\varepsilon\xspace}
\newcommand{\fE}{E\xspace}
\newcommand{\HD}{HD\xspace}
\newcommand{\HDs}{HDs\xspace}

\newcommand{\mf}{\mathscr{F}}
\begin{document}

\title{Emergence of Quasi-equilibrium State and Energy Distribution \\ for the Beads-spring Molecule Interacting with a Solvent}
\author{Tatsuo Yanagita}
\affiliation{Department of Engineering Science, Osaka Electro-Communication University, Neyagawa 572-8530, Japan}
\email{yanagita@osakac.ac.jp}

\author{Tetsuro Konishi}
\affiliation{General Education Division, College of Engineering, Chubu University, Kasugai 487-8501, Japan}
\email{tkonishi@isc.chubu.ac.jp}

\begin{abstract}
We study the energy distribution during the emergence of a quasi-equilibrium (\QE) state  in the course of relaxation to equipartition in slow-fast Hamiltonian systems. 
A bead-spring model where beads (masses) are connected by springs is considered, and it is used as a model of polymers.
The \QE lasts for a long time because the energy exchange between the high-frequency vibrational and other motions is prevented when springs in the molecule become stiff.
We numerically calculated the time-averaged kinetic energy and found that the kinetic energy of the solvent particles was always higher than that of the bead in a molecule.
This is explained by adapting the equipartition theorem in \QE, and it agrees well with the numerical results. 
The energy difference can help determine how far the system is from achieving equilibrium, and it can be used as an indicator of the number of frozen or inactive degrees exist in the molecule.
\end{abstract}
\date{\today }

\maketitle

\section{Introduction}
\label{sec:intro}

Relaxation to equipartition in Hamiltonian dynamical systems is a long-standing problem that has been extensively studied; for example, in the coupled oscillatory chain, i.e., the Fermi--Pasta--Ulam systems \cite{FPU1965}.
Systems considered in these studies are nearly integrable, and the relaxation to equipartition is often prevented by the KAM tori.
Thus, there is an energy threshold to determine whether an equipartition state is established \cite{PhysRevA.31.1039}; further the relaxation shows multiple stages and non-monotonic behavior \cite{PhysRevE.92.022917}.
The energy transfer between fast and slow subsystems becomes very slow when the difference between timescales in the subsystems is very large.
This effect was fast noticed by the Boltzmann and Jeans \cite{boltzmann-1895}, and later, Landau and Teller presented the exponential law of relaxation time \cite{Jeans-1903,Jeans-1905,Jeans-93}.
This effect has been confirmed numerically for a classical gas of diatomic molecules \cite{jeans-numeric-91}; further, such systems are known to prevent equipartition  \cite{benettin-jeans-2,benettin-jeans-ptp-94,benettin-jeans-nonlin-96}.
In addition, a relaxation process is considered in the Hamiltonian dynamics of self-gravitating systems \cite{sheet-tkg-1,sheet-tgk-2,sheet-tgk-3}.
These systems demonstrate the existence of quasi-equilibrium and slow relaxation to the equilibrium.
Further, in the context of the function of proteins, the energy transfer occurring out-of-equilibrium plays a key role, and there exists a bottleneck for this energy transfer \cite{doi:10.1021/acs.jpclett.5b02514}.

We consider the dynamics of the bead-spring model--- masses connected by springs---that can be considered as models of protein and DNA and have practical importance and play  significant roles in various fields \cite{doi-edwards}.  
In terms of the function of these materials, it is important to consider the dynamic activity during relaxation to equilibrium.

In our previous studies, we focused on the relaxation process in a serially connected bead-spring molecule.
When the molecule is surrounded by a solvent, the beads exchange energy by collisions with solvent particles, and the system relaxes to equipartition, where all kinetic energies of the beads in the molecule and solvent particles become equal.
We regard the kinetic energy of each component as an indicator of dynamical activeness.
Further, we reported tha the distribution of the time-averaged energy of each bead is not uniform over a long time, which we call the quasi-equilibrium (\QE)  \cite{konishi-yanagita-chain, konishi-yanagita-solvent}.
 If the spring constant $k$ of the spring between the beads in the molecule is large, the duration of \QE obeys $\exp(c\sqrt{k})$, which is a typical feature of the Boltzmann-Jeans type relaxation \cite{konishi-yanagita-solvent}. 

In this study, we consider the kinetic energy per bead in the molecule and the solvent particle in the \QE.
We consider two types of molecules: chain and network; these are respectively sparsely and densely connected.
We numerically showed that the time-averaged kinetic energy of the solvent particles was larger than that of the beads in the bead-spring molecule in \QE.
The difference between the time-averaged kinetic energy of the molecule and the solvent depends on the type of molecule (the connection topology), the size of the molecule, and the number of solvent particles.
We theoretically clarify that the difference in the kinetic energies by adapting the equipartition theorem.
The key point of the emergence of the energy difference in \QE is the existence of ``frozen'' degrees, which comes from high-frequency vibrations in the molecule.
These vibrational motions are frozen in the sense that the energy exchange to the other part is prevented.
We estimate the number of frozen degrees and adapt the equipartition theorem to the \QE state, and we analytically determine the extent to which the kinetic energy of the solvent particle is higher than that of the molecule.
The theoretical analysis is in good agreement with the numerical results.




The rest of this manuscript is organized as follows:
In Sec.~\ref{sec:model}, we introduce a system of bead-spring molecules in a solvent, and
we present the Hamiltonian of the models.
The numerical method and parameter settings are discussed in Sec.~\ref{sec:method}.
The numerical results are shown in Sec.~\ref{sec:numerics}.
In Subsec.~\ref{subsec:qstate}, the slow relaxation to equipartition and the emergence of \QE are shown where the kinetic energy of the solvent particle is higher than that of bead in the molecule.
We show the kinetic energy difference between bead in molecule and solvent particles in Subsec.~\ref{subsec:chain_qstate}.
Numerical results for the network molecules, which are beads are connected by springs with a network topology as a model of a densely-connected molecule, are presented in  Subsection~\ref{subsec:network}.
The theoretical explanation for the numerical results is provided based on the equipartition theorem in Sec.~\ref{sec:theory}.
Finally, Sec.\ref{sec:summary} summarizes this work and provides a brief outlook.

\section{Bead-spring molecule in solvent}
\label{sec:model}

The bead-spring molecule considered in this study is one where the masses are connected by springs.
The connection can be a serial or complex network structure.
These bead-spring molecules are in the external potential, and they interact with the solvent particles.
For simplicity, we consider that the beads are connected by linear springs.
Even with the linear spring used in this study, the system is nonlinear and exhibits strong chaotic behavior.
The schematic of the models for the serial and network connections are shown in Fig.~\ref{fig:model1} and \ref{fig:model2}, respectively.

The Hamiltonian of the system is given by
\begin{align}
H &= H_{\rm mol} + H_{\rm sol} + H_{\rm int}, 
\label{eq:hamiltonian-all}\\
H_{\rm mol} & = \sum_{i=1}^{N_{b}} \frac{p_i^2}{2m} 
+ \sum_{i\neq j}^{N_{b}}\frac{k_{i,j}}{2}\left(\left| \VEC{r_{j}}-\VEC{r_i}\right| -\ell_{i,j}\right)^2 \nonumber \\
&+ \sum_{i=1}^{N_{b}} U_{\rm ext}\left(\VEC{r_i}\right)\ ,
 \label{eq:hamiltonian-chain}\\
H_{\rm sol} &= \sum_{\alpha=1}^{N_{s}} \frac{p_\alpha^2}{2m} +
\sum_{\alpha=1}^{N_{s}} U_{\rm ext}\left(\VEC{r_\alpha}\right) \ , 
\label{eq:hamiltonian-solvent}\\
H_{\rm int} &= \sum_{i,\alpha;{\rm pair}} U_{\rm int}\left(\left|\VEC{r_i}-\VEC{r_\alpha}\right|\right) \nonumber\\
&+ \sum_{\alpha,\beta; {\rm pair}}U_{\rm int}\left(\left|\VEC{r_{\alpha}}-\VEC{r_{\beta}}\right| \right)\ , 
\label{eq:hamiltonian-interaction} 
\end{align}
where $N_b$ and $N_s$ denote the number of beads in the molecule and solvent particles, respectively, and the subscripts $i$ and $j$ represent the indices of beads in the molecule, and $\alpha$ and $\beta$ represent the indices of solvent particles. 
Here, $\VEC{r_i}$ and $\VEC{r_\alpha}$ represent the positions of beads $i$ and solvent particles $\alpha$, respectively, where
 bead $i$ belongs to the molecule and particle $\alpha$ belongs to the solvent, while
$\VEC{p_i}$ and $\VEC{p_\alpha}$ are the momenta conjugate to 
$\VEC{r_i}$  and $\VEC{r_\alpha}$, respectively. 
The natural length of the spring between $i$ and $j$ beads is $\ell_{i,j}$.
Solvent particles interact with beads in the molecule via the potential
$U_{\rm int}$
\begin{equation}
  \label{eq:potential-int}
  U_{\rm int}(r)= 
  \begin{cases}
\displaystyle
\frac{k_{\alpha}}{2}(r-l_\alpha)^2 
& \cdots r < l_\alpha  \ , \\
0 & \cdots r \ge l_\alpha \ .
  \end{cases}
\end{equation}

The external potential $U_{\rm ext}$ confines the molecule and solvent in a 
finite region via
\begin{equation}
U_{\rm ext}(\VEC{r})=a \sum_{n=1}^{N_{\rm wall}} \left|\left|\VEC{r} - \VEC{R}_n \right|-b\right|^{-6}.
\label{eq:potential-external} 
\end{equation}
It breaks the rotational symmetry and thus prevents the conservation of the angular momentum.
Here, we use the following parameters: $a=0.01, N_{\rm wall}=4$, $b=4N\ell$, 
$\vec{R_j}\equiv (R,0)$, $ (0,R)$, $ (-R,0)$, $(0,-R)$, and
$R\equiv N\ell + \sqrt{a^2-N^2\ell^2}$.
This potential resembles the one used in dispersing billiards~\cite{billiards-book},
where systems do not have any conserved quantities other than 
the total energy, and the orbit  obtained from long time simulation
can be well approximated by the micro-canonical distribution.

We consider two types of bead-spring molecules in two-dimensional space: bead-spring chain and bead-spring network molecules.
These models correspond to the two cases, in which the beads in the molecule are sparsely and densely connected by springs, respectively.

\subsection{Bead-spring chain molecule}
\label{subsec:model_chain}

\begin{figure}[tb]
\begin{center}
\resizebox{\singlefiguresize}{!}{\includegraphics{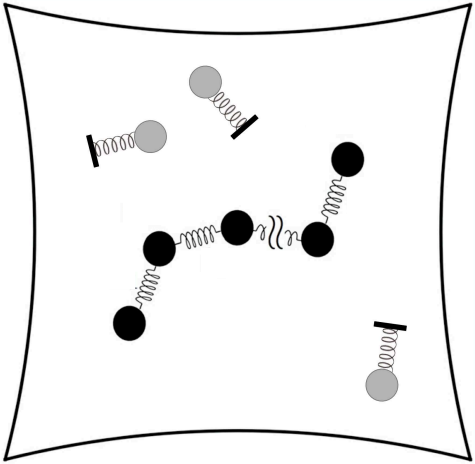}}
\end{center}
\vspace{-10mm}
\caption{
Schematic of the bead-spring chain molecule in solvent particles. 
The beads are serially connected by linear springs.
The figure shows the chain molecule with $N_{b}=N$ and the number of solvent particles $N_{s}=3$.
}
\label{fig:model1}
\end{figure}

The chain-type molecule is that beads are connected serially with springs (Fig.~\ref{fig:model1}).
The following homogeneous parameters are used here
\begin{equation}
  k_{i,j}= 
  \begin{cases}
\displaystyle
k
& (i=1,\cdots, N_{b}-1,\ j=i+1) \ , \\
0 &{\rm otherwise}
  \end{cases}
\end{equation}
and $\ell_{i,j}=1 \ (i=1,\cdots,N_{b}-1,\ j=i+1)$.

For the initial condition of the chain molecule, all springs between beads were set to be the natural length, and
the initial positions of beads $\VEC{r_i}(0)=(x_i(0),y_i(0))$ are
\begin{equation}
  \label{eq:initial-position-chain}
x_i(0) = i-(N+1)/2, \  y_i(0) = 0\ \ (i=1,\cdots,N_{b}),
\end{equation}
respectively, which is the center of the mass that sets to be the origin $\vec{r}_{\rm CM}(0)=\sum_{i}m_i\vec{r_i}(0)/\sum_i m_i=(0,0)$.
In addition, the initial velocity of the chain molecule sets 
\begin{align}
  \label{eq:initial-velocity-chain}
\dot{x_i}(0) &=v_0 \cos(\theta_0),\nonumber\\
\dot{y_i}(0) &= v_0\sin(\theta_0) \ \ (i=1,\cdots,N_{b}),
\end{align}
where the initial direction of the motion $\theta_0$ is a random number selected from a uniform distribution in the interval $[0,2\pi]$.
Thus, the total energy of the chain molecule is $K_{\rm chain}=\frac{M}{2} v_0^2$, where $M=mN_{b}$ is the total mass.


\subsection{Bead-spring network molecule}
\label{subsec:model2}

\begin{figure}[tb]
\begin{center}
\resizebox{\singlefiguresize}{!}{\includegraphics{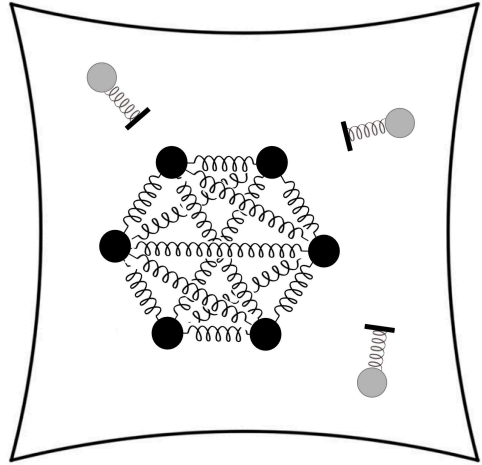}}
\end{center}
\vspace{-10mm}
\caption{
Schematic of the bead-spring network molecule. 
The beads are densely connected with linear springs.
The connection topology considered here is complete graph.
The figure shows the network molecule with $N_{b}=6$ and the number of solvent particles $N_{s}=3$.
}
\label{fig:model2}
\end{figure}

As shown below, the number of springs is a crucial parameter that determines the energy distribution during \QE.
Thus, it is interesting to consider the bead-spring network molecule where the connections of beads in the molecule are represented by the adjacency matrix $k_{i,j}$.
To contrast the chain molecule, we consider the complete graph as the connection topology of the beads in the molecule
where all pairs of beads in the molecule are connected by springs.
The Hamiltonian of the network molecule is 
\begin{align}
H &= H_{\rm net} + H_{\rm sol} + H_{\rm int}, \label{eq:hamiltonian-net1}\\
H_{\rm net} & = \sum_{i=1}^{N_{b}} \frac{p_i^2}{2m} \nonumber\\
&+ \sum_{i=1}^{N_{b}-1} \sum_{j = i+1}^{N_{b}}\frac{k_{i,j}}{2}\left(\left| \VEC{r_{j}}-\VEC{r_i}\right| -\ell_{i,j}\right)^2.
 \label{eq:hamiltonian-net2}
\end{align}
The Hamiltonians of solvent $H_{\rm sol}$ and the interaction parts $H_{\rm int}$  are the same as in Eqs.~(\ref{eq:hamiltonian-solvent}) and (\ref{eq:hamiltonian-interaction}).
The external potential $U_{\rm ext}$ is the same as in Eq.~(\ref{eq:potential-external}). 
A schematic of the network molecules is shown in Fig.~\ref{fig:model2}.

For the initial condition, we consider that each bead in the molecule is located at the vertex of the $N_b$ equilateral polygon placed on the unit circle.
In particular, the initial positions of the beads $\VEC{r_i}(0)=(x_i(0),y_i(0))$ are
\begin{align}
  \label{eq:initial-position-net}
x_i(0)&=\cos(2 \pi  i/N_{b}), \nonumber\\
y_i(0) &= \sin(2 \pi i/N_{b}) \ \ (i=1,\cdots,N_{b}),
\end{align}
respectively.
We set the natural lengths of springs to $\ell_{i,j}^2=(x_i(0)-x_j(0))^2+(y_i(0)-y_j(0))^2$.
The natural length of the spring is not the same value for the network molecule, and all lengths of the spring are set to be the natural length initially; thus, the potential part of the Hamiltonian Eq.~(\ref{eq:hamiltonian-chain}) is zero.

\section{Numerical method and parameter settings}
\label{sec:method}

For time integration, we use the fourth-order implicit  Runge--Kutta method with a time interval $\delta t=0.001/\sqrt{k}$.
The method is known to be symplectic, and the total energy is well conserved in our numerical simulations.

Throughout the paper, all masses are the same, i.e.,
\begin{equation}
m_{\alpha}=m_i = m = 1 \ \mbox {for all} \ i \mbox{ and } \alpha.
\end{equation}
Further, we set that all spring coefficients between beads in the molecule are same 
\begin{equation}
k_{i,j}=k  \mbox{ for } i,j=1,\cdots,N_{b}.
\end{equation}

The initial positions of the solvent particles randomly set with the distance of any pair of particles
is larger than $l_{\alpha}=1$; $|\VEC{r_{\alpha}}-\VEC{r_{\beta}}|>\ell_{\alpha}$ and $|\VEC{r_{i}}-\VEC{r_{\alpha}}|>\ell_{\alpha}$ for all $\alpha$ and $\beta$, and $i$ and $\alpha$, respectively. 
Therefor, the interaction energy between the beads and solvent particles is initially zero, i.e., $H_{\rm int}=0$.
For solvent particles, the kinetic energy was set to be zero $p_{\alpha}=0$ for all $\alpha$.
All momenta of the beads in the molecule are set to the same value, which means that the molecule has only the kinetic energy of the center of gravity.
We can say that the above initial condition is a scenario in which a ``hot'' molecule is set in a ``cold'' solvent.
In other words, the system is highly inhomogeneous in the sense that all energy is placed into the kinetic part of the molecular beads, and the other parts of the energy are set to zero.

As we are interested in the energy distribution during the \QE state, we focus on the scenario wherein the fast timescale come from the high-vibrational motion of the molecule, which comes from the large spring constant, and we set $k=1000$. For the interaction between the molecule and solvent, we set $k_{\alpha}=1$ in Eq.~(\ref{eq:potential-int}).

\section{Numerical results}
\label{sec:numerics}

To observe the relaxation process, we measured the time-averaged kinetic energy ${K_i}(t)$ of the beads and solvent particles, and the spring energies ${V_{i,j}}(t)$ in the molecule, which are defined as 
\begin{equation}
   \label{eq:average-kinetic-energy}
   {K_i}(t)\equiv \frac{1}{t}\int_0^t \frac{p_i^2(t')}{2m} \,dt' ,
\end{equation}
\begin{equation}
   \label{eq:average-spring-energy}
   {V_{i,j}}(t)\equiv \frac{1}{t}\int_0^t \frac{k_{i,j}}{2}\left(\left| \VEC{r_{j}}(t')-\VEC{r_i}(t')\right| -\ell_{i,j}\right)^2 \,dt' .
\end{equation}
We averaged the time-averaged kinetic energy over all beads and the particles as 
\begin{eqnarray}
\overline{K_{\rm mol}}(t)&=&\frac{1}{N_{b}}\sum_{i=1}^{N_{b}}  {K_i}(t)  \\
\overline{K_{\rm sol}}(t)&=&\frac{1}{N_{s}}\sum_{\alpha=1}^{N_{s}}  {K_\alpha}(t)  \\
\overline{V_{\rm mol}}(t)&=&\frac{1}{N_{k}} \sum_{i\neq j}^{N_{b}}{V_{i,j}}(t),
\end{eqnarray}
where  $N_{k}=\sum_{i \neq j} \Theta(k_{i,j})$ is the number of springs in the molecule, and $\Theta(x)$ is the Heaviside function. 
Because the time evolution of the above kinetic energies depends on the initial condition, the initial moment of the molecule, and the initial position of the solvent particles, we further averaged it over $M$ ensembles starting from different initial conditions.

\begin{figure}[tb]
\begin{center}
\resizebox{\singlefiguresize}{!}{\includegraphics{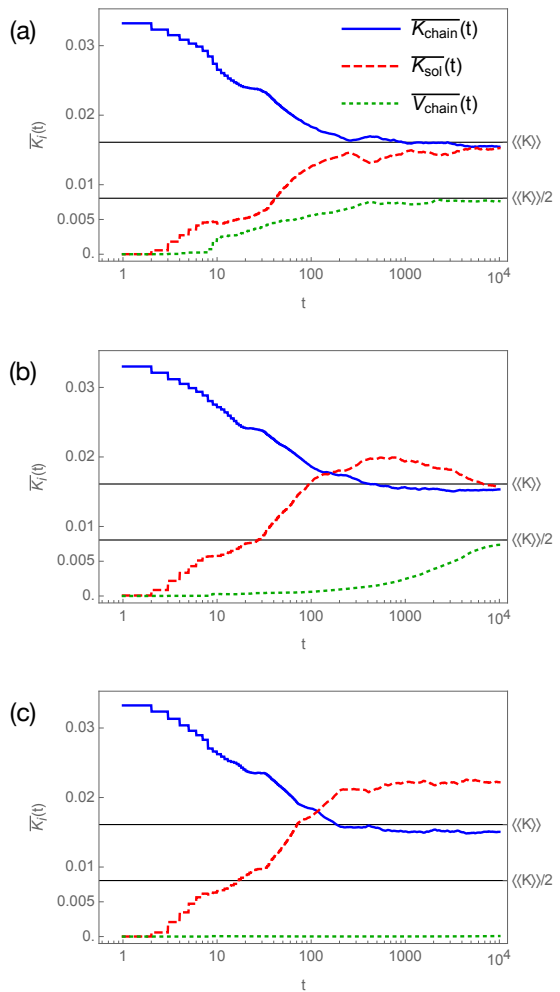}}
\end{center}
\vspace{-5mm}
\caption{
Relaxation process to equipartition.
Kinetic energy of chain molecules and solvent particles as a function of the averaging time.
The solid, dashed, and dotted curves represent the kinetic energies of the chain molecule and solvent particles, and the potential energy of the springs, respectively.
The horizontal solid lines represent the estimated energy level when the equipartition is established.
Spring constants between chain beads are (a) $k=1$,  (b) $k=10$, and (c) $k=100$.
The ensemble average is taken over 20 different initial conditions.
$N_{b}=3$ , $N_{s}=2$, $\ell_{i,j}=\ell_{\alpha}=1$,  $m=1$, $k_{\alpha}=1$, and $E_0=0.1$.
}
\label{fig:qstate}
\end{figure}

\begin{figure}[tb]
\begin{center}
\resizebox{\singlefiguresize}{!}{\includegraphics{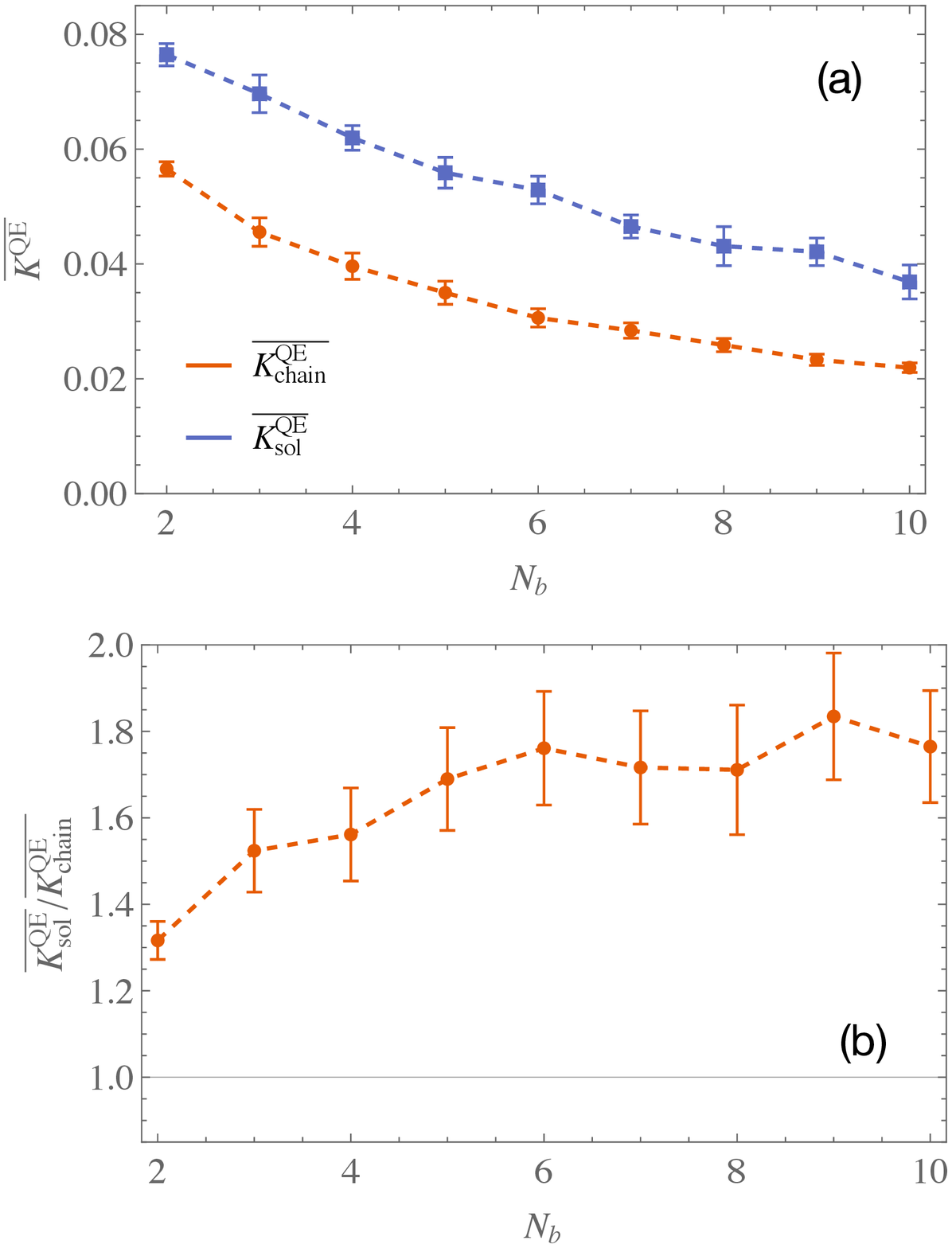}}
\end{center}
\vspace{-5mm}
\caption{
(a) Time-averaged kinetic-energies of the chain molecule $\tav{K_{\rm chain}}$ and the solvent  $\tav{K_{\rm sol}}$ as a function of chain length $N_{b}$.
The circles and the squares with the error bars are $\tav{K_{\rm chain}}$ and $\tav{K_{\rm sol}}$, respectively.
(b) The ratio $\tav{K_{\rm mol}}/\tav{K_{\rm sol}}$ is shown as a function of $N_{b}$.
$k=1000$, $N_{s}=2$, $k=1000$, $M=10$, $T_\quasi=10^4$ and $E_0=0.3$. 
}
\label{fig:chain_ncdep}
\end{figure}
\begin{figure}[tb]
\begin{center}
\resizebox{\singlefiguresize}{!}{\includegraphics{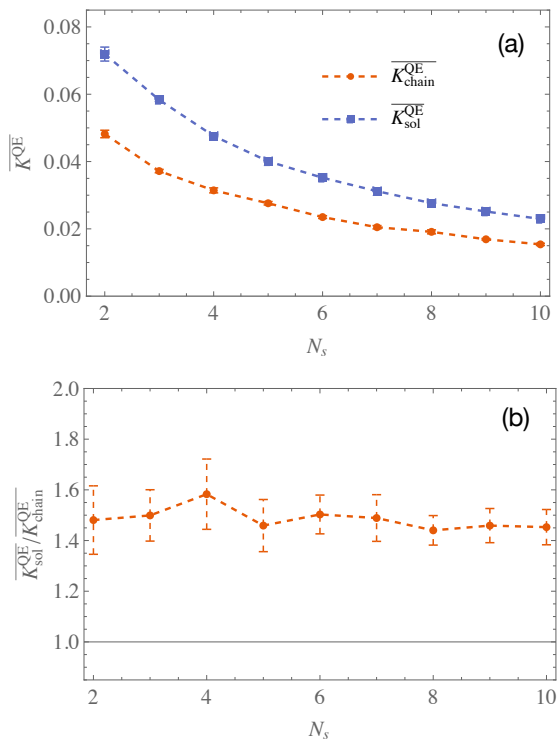}}
\end{center}
\vspace{-5mm}
\caption{
(a) Time-averaged kinetic energies of the chain molecules $\tav{K_{\rm chain}}$ (circles) and the solvent particles $\tav{K_{\rm sol}}$ (squares) in \QE as a function of the number of solvent particles $N_{s}$.
(b) The ratio $\tav{K_{\rm sol}}/\tav{K_{\rm chain}}$  dependency on $N_{s}$.
The other parameters are the same as those in Fig.~\ref{fig:chain_ncdep}.
}
\label{fig:chain_nsdep}
\end{figure}

\subsection{Slow relaxation to equipartition and the emergence of quasi-equilibrium}
\label{subsec:qstate}
We demonstrate the relaxation to equipartition through the short-chain molecule with the number of beads $N_{b}=3$ surrounded by two solvent particles $N_{s}=2$.
The evolution of the time-averaged kinetic energies of the chain molecule and solvent, and the potential energy of the spring are shown in Fig.~\ref{fig:qstate}.
When the two spring constants are equal, $k=k_{\alpha}=1$, there are no fast slow-time scales in the system.
The initially assigned kinetic energy to the chain molecule is distributed gradually into  other parts such as the vibrational and interaction parts.
As shown in Fig.~\ref{fig:qstate}(a), the system relaxes to the equipartition wherein all kinetic energies for the beads and solvent particles are equal.
The solid curve shows the kinetic energy of the chain molecule $\overline{K_{\rm mol}}(t)$, and it is almost monotonically decreases and converges to a stationary value.
As shown in the dashed curve, the kinetic energy of solvent $\overline{K_{\rm sol}}(t)$  monotonically increases from zero and converges to the same stationary value as that of the chain molecule.
The spring energy $\overline{V_{\rm mol}}(t)$ shown by the dotted line, gradually increases from zero and converges to a stationary value, that is, half of the stationary value of the time-averaged kinetic energy because we consider two-dimensional space.
\[
\lim_{t \rightarrow \infty}\overline{V_{\rm mol}}(t)=\lim_{t \rightarrow \infty}\frac{\overline{K_{\rm mol}}(t)}{2}
\]

However, the relaxation time to equipartition depends on the gap between the timescales involved in the system.
In our system, two timescales are involved, i.e., the time scale of the vibration of the molecule and that of the interaction between the molecular bead and the solvent particle.
These timescales are determined by the spring constants $k$ and $k_{\alpha}$.
When the difference between these time scales is larger, the relaxation to the equipartition takes a long time.
Furthermore, the relaxation process is not monotonic; rather, the time-averaged kinetic energy of the solvent particle shows an overshoot.
In fact, as the spring constant $k$ increases, we observe that the kinetic energy of the solvent starting from zero increases gradually, and it exceeds that of the chain molecule, as shown in Fig.\ref{fig:qstate}(b).
After some period, it gradually relaxes to the equipartitioned value.
For the potential energy of the spring in the molecule, we see that $\overline{V_{\rm mol}}(t)$ increases gradually, and it converges to half of the equipartitioned value.

As the spring constant $k$ becomes considerably larger in addition to the longer relaxation time, a plateau appears as shown in Fig.\ref{fig:qstate}(c) after $\overline{K_{\rm sol}}(t)$ exceeds $\overline{K_{\rm mol}}(t)$.
Because the plateau becomes wider as $k$ increases, the system is almost stationary.
We call this stationary state in which the system settles in the plateau a quasi-equilibrium (QE).
After a long \QE, the system relaxes to the equipartition.
The duration of \QE becomes longer with $k$.
Moreover, the duration increases as $\exp(\sqrt{k})$, which is consistent with the Boltzmann--Jeans conjecture, as reported in \cite{konishi-yanagita-chain,konishi-yanagita-solvent}.
Hereafter, we clarify how the kinetic energies of the bead molecule and solvent particles are distributed during the \QE.

\begin{figure}[tb]
\begin{center}
\resizebox{\singlefiguresize}{!}{\includegraphics{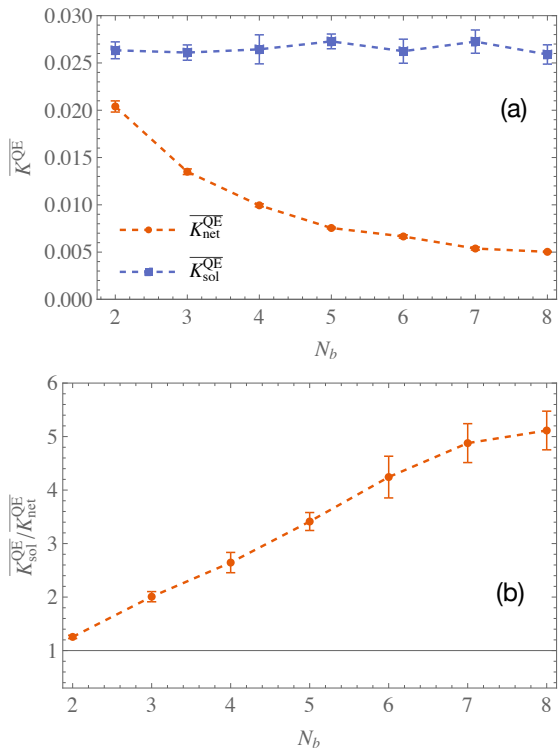}}
\end{center}
\vspace{-5mm}
\caption{
(a) Time averaged Kinetic energies of the beads in the network molecules and those of the solvent particles are shown as a function of the number of beads by the circle and the square with error bars, respectively.
(b) The ratio of the kinetic energy of the solvent particle to that of the network molecule is shown as a function of the number of beads.
$N_{b}=6$ and $E_0=0.1$.
The other parameters are the same as those in Fig.~\ref{fig:chain_ncdep}
}
\label{fig:cgraf_nbdep}
\end{figure}

\begin{figure}[tb]
\begin{center}
\resizebox{\singlefiguresize}{!}{\includegraphics{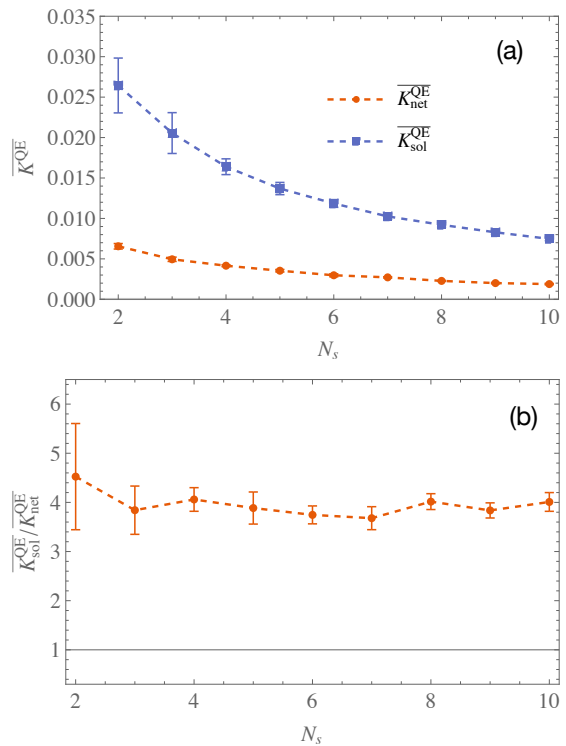}}
\end{center}
\vspace{-5mm}
\caption{
(a) Time-averaged kinetic energies of the beads in the network molecules and solvent particles are shown by the circles and the squares with error bars, respectively.
(b) The ratio of the kinetic energy of the solvent to that of the network molecule is shown as a function of the number of solvent particles $N_{s}$. 
The ratio does not depend on $N_{s}$.
The other parameters are the same as those in Fig.~\ref{fig:cgraf_nbdep}
}
\label{fig:cgraf_nsdep}
\end{figure}

The \QE appears for both systems of the chain and network molecules, and the existence of \QE is irrelevant to the connection structure of the molecule.
During the \QE, we found that the universal feature that the average kinetic energy of solvent particles was larger than that of the molecular beads (see Fig~\ref{fig:qstate}).
To characterize this difference in \QE, we observe the kinetic energy of each particle at which the \QE state persists, i.e., $\overline{K_{\rm mol}^\quasi}=\overline{K_{\rm mol}}(t_\quasi)$ and $\overline{K_{\rm sol}^\quasi}=\overline{K_{\rm sol}}(t_\quasi)$, where $t_\quasi$ is a typical time during the \QE state, and we use $t_\quasi=10^4$. For \QE , we find that the following inequality holds.
\[
\overline{K_{\rm mol}}(t_{\quasi})<\overline{K_{\rm sol}}(t_{\quasi}).
\]


In the following subsections, we consider the time-averaged kinetic energy distribution for molecular beads and solvent particles in \QE by changing the molecular size and structure and the number of solvent particles.

\subsection{Quasi-equilibrium for the chain molecule}
\label{subsec:chain_qstate}

The kinetic energies $\overline{K_{\rm chain}^\quasi}$ and $\overline{K_{\rm sol}^\quasi}$
 that depend on the number of beads in a chain molecule are shown in Fig.~\ref{fig:chain_ncdep}.
Both energies decrease gradually, and the difference between them converges to a constant value as $N_{b}$, i.e., the length of the chain molecule, increases.
It is stressed that, $\overline{K_{\rm mol}}(t_{\quasi})<\overline{K_{\rm sol}}(t_{\quasi})$ holds for any chain length.
Indeed, as shown in Fig.~\ref{fig:chain_ncdep}(b), the ratio of the kinetic energy of the chain molecule to that of the solvent converges to a value as $N_{b}$ increases.
The ratio is always larger than one, and it increases gradually as $N_{b}$ increases.
This clearly shows that the kinetic energy of the solvent is always larger than that of the chain molecules during \QE.


Fig.~\ref{fig:chain_nsdep}(a) shows the $\overline{K_{\rm chain}^\quasi}$ and $\overline{K_{\rm sol}^\quasi}$ dependency on the number of solvent particles $N_{s}$.
Both energies gradually decrease as $N_{s}$.
However, the ratio $\overline{K_{\rm sol}^\quasi}/\overline{K_{\rm chain}^\quasi}$ does not depend on the number of solvent particles, as shown in Fig.~\ref{fig:chain_nsdep}(b).
These results support that in \QE, the kinetic energy of the solvent particles is always larger than that of the chain molecule.
It is stressed that the ratio is independent of the number of solvent particles, which is usually large in reality.

\subsection{Quasi-equilibrium for the network molecule}
\label{subsec:network}
We consider the energy distribution during \QE for the network molecule where the beads in the molecule are densely connected by linear springs as shown in Fig.\ref{fig:model2}.
To contrast the effect of the number of springs which plays a crucial role in energy distribution, we consider the complete graph as the connection topology for the network molecule.
The Hamiltonian is defined by Eqs.~(\ref{eq:hamiltonian-net1}) and (\ref{eq:hamiltonian-net2}).
In Fig.~\ref{fig:cgraf_nbdep}(a), the time-averaged kinetic energy for the network molecule, and that for solvent particles $\overline{K_{\rm net}^\quasi}$ and $\overline{K_{\rm sol}^\quasi}$ are shown as a function of $N_{b}$, i.e., the number of beads in the network molecule.
While $\tav{K_{\rm net}}$ gradually decreases with $N_{b}$, $\tav{K_{\rm sol}}$ does not depend on $N_{b}$.
In Fig.~\ref{fig:cgraf_nbdep}(b), the ratio $\tav{K_{\rm sol}}/\tav{K_{\rm net}}$ is shown as a function of $N_{b}$.
In contrast to the chain molecule, the ratio increases linearly with $N_{b}$ and is always larger than one.

In this section, we numerically show that the kinetic energy of the solvent particle is always larger than that of the molecular bead in \QE, i.e., $\overline{K_{\rm mol}}(t_{\quasi})<\overline{K_{\rm sol}}(t_{\quasi})$. 
This fact is independent of the molecular structure, the number of beads in the molecule, and that of the solvent particles.
In other words, the solvent particles are more energetic than the beads in the molecule during the \QE. 
The number of springs in the molecule plays a crucial role in determining the distribution of energy.

\section{Theoretical explanation}
\label{sec:theory}

\subsection{Equipartition theorem}
\label{subsec:equi}

Let us recall the equipartition theorem.
Usually, the theorem states that the thermal average of the kinetic energy in the equilibrium is given by
\begin{equation}
\cav{\fe}\equiv
\cav{\frac{(p_i^x)^2}{2m}}
=\cav{\frac{(p_i^y)^2}{2m}}
=\frac{1}{2\beta},
\end{equation}
where $p_i^x$ and $p_i^y$ are the momenta of the $x$ and $y$ directions of particle $i$ in the system, respectively.
The symbol $\cav{\cdots}$ represents the thermal average at the inverse temperature
$\beta\equiv 1/k_B T$, and it is defined as
\begin{equation}
\cav{f(q,p)}\equiv
\frac{1}{Z}\int f(q,p) e^{-\beta H(p,q)}d\Gamma 
 \label{eq:thermal-ave-def}
\end{equation}
for any function of general coordinates and their conjugate momenta $f(q,p)$.
Here, $d\Gamma$ denotes a volume element of the phase space, and $Z$ represents a partition function.
The key point of the theorem is that the kinetic energy is harmonic in the momentum.
Further, if any energy described by the harmonic form of the coordinate $q$, we can show that the thermal average of such potential energy also takes the same value \cite{kubo-book}.
For example, the thermal average of the potential energy with in form $U(q)=\frac{1}{2} q^2$ is
\begin{equation}
\cav{U(q)} =\frac{1}{2 \beta}=\cav{\fe}.
\end{equation}
Consider the following Hamiltonian representing by the sum of the harmonic form
$$
\mathscr{H}=\sum_{i=1}^{f} \frac{(p_i^x)^2+(p_i^y)^2}{2m_i}+\sum_{i=1}^{f}\frac{1}{2}{k_i}q_i^2,
$$
where $2f$ is the number of degrees of freedom.
Then, the thermal average of the energy of such a system becomes
\begin{equation}
\cav{\mathscr{H}}
=\frac{\mf_{\rm sys}}{2\beta}=\mf_{\rm sys}\cav{\fe},
\end{equation}
where we define $\mf_{\rm sys}$ as the number of independent harmonic terms that enter into the Hamiltonian.
For the above Hamiltonian, the total number of harmonic degrees is $\mf_{\rm sys}=3f$.
\begin{figure}[tb]
\begin{center}
\resizebox{\singlefiguresize}{!}{\includegraphics{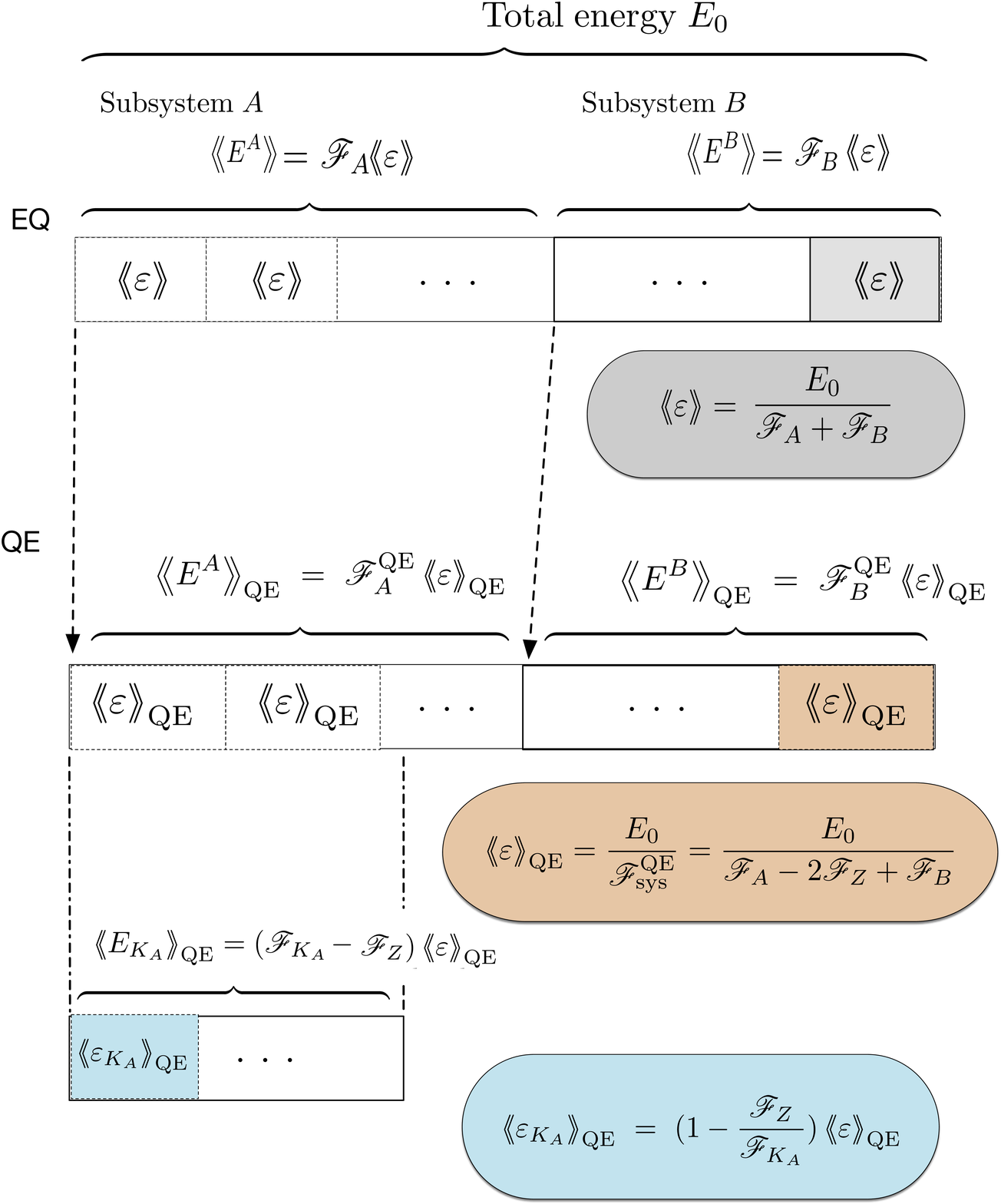}}
\end{center}
\vspace{-5mm}
\caption{
Schematic for the energy distribution per \HD based on the equipartition theorem.
The top panel shows the equipartition state in equilibrium, where the total energy $E_0$ is equally distributed to the total number of \HDs, $\mf_{\rm sys}=\mf_{A}+\mf_{B}$.
The middle panel shows the \QE state at which the total energy is equally distributed to ``active'' \HDs, i.e., $\mf_{\rm sys}^\quasi=\mf_{A}-2\mf_{Z}+\mf_{B}$.
The bottom panel shows that the kinetic energy of subsystem $A$ is redistributed to the number of kinetic \HDs, that is, $\mf_{K_A}$.
}
\label{fig:schematic-equipartition}
\end{figure}
\subsection{Theoretical estimation of averaged kinetic energies for quasi-equilibrium}
\label{subsec:equipartition-qstate}

We now estimate the kinetic energy of the molecular beads and solvent particles by adapting the equipartition theorem for QE.
To this end, we assume the following.
First, we assume that the thermal average of the total energy $\cav{\mathscr{H}}$ is equal to $E_0$.
Second, anharmonic potential terms are approximated by harmonic ones.
The second assumption is valid for a scenario wherein the amplitude of the vibrational oscillation is sufficiently small because the spring coefficient is sufficiently large.
%

Based on these assumptions, the number of independent harmonic terms entering the Hamiltonian is denoted by $\mf_{\rm sys}$; we refer to them as the harmonic degrees (\HDs).
Thus, we consider that the total energy $E_0$ is equally distributed into $\mf_{\rm sys}$ \HDs.
Therefore, in equilibrium, the energy per \HD can be estimated by
\begin{equation}
\eav{\fe}\equiv\frac{E_0}{\mf_{\rm sys}}.
\end{equation}
Hereafter, we represent $\eav{\cdots}$ as the kinetic energy per \HD by adapting the equipartition theorem to the Hamiltonian dynamical system.

To estimate the energy per \HD, we divide the system into two subsystems $A$ and $B$, where the subsystem $A$ has frozen \HDs, and $B$ does not have such frozen degrees.
Furthermore, we denote the number of \HDs in equilibrium for $A$ and $B$ by $\mf_{A}$ and $\mf_{B}$, respectively.
Thus, the total number of \HDs of the system is $\mf_{\rm sys}=\mf_{A}+\mf_{B}$.
When we adapt the equipartition in the equilibrium, the equally distributed energy per \HD can be estimated by
\begin{equation}
\eav{\fe}
=\eav{\fe^A }
=\eav{\fe^B }
=\frac{E_0}{\mf_{\rm sys}}=\frac{E_0}{\mf_{A}+\mf_{B}}.
\end{equation}

A schematic of the energy distribution in the equilibrium is shown in the top panel of Fig.~\ref{fig:schematic-equipartition}.
In the case of \QE, the time-averaged spring energy is almost zero if the spring coefficient is sufficiently large, as shown in Fig.~\ref{fig:qstate}(c).
This means that the vibrational motion of the molecule is frozen in the sense that there are no energy exchanges between them and the other degrees.
To estimate the energy per \HD in \QE, we describe the number of \HDs of subsystem $A$ as the sum of the number of kinetic \HDs and that of the potential \HDs, i.e., 
$\mf_{A}=\mf_{K_A}+\mf_{V_A}$, where $\mf_{K_A}$ denotes the number of \HDs for the kinetic part, and $\mf_{V_A}$ denotes the number of \HDs for the potential part entered in the Hamiltonian of subsystem $A$, respectively.
When the number of stiff springs in $A$ is $\mf_{Z}$, the number of active \HDs in \QE is $\mf_{A}^\quasi
=\mf_{A}-2\mf_{Z}$.
Here, the factor two originates from the following fact that these stiff springs behave as rigid links, and they act as constraints for molecular motion.
Thus, the number of \HDs for the kinetic part is reduced to $\mf_{K_A}-\mf_{Z}$, and that of the \HDs for the potential part is reduced to $\mf_{V_A}-\mf_{Z}$.
The bead-spring system in \QE behaves as a system with constraints in equilibrium \cite{konishi-yanagita-chain, konishi-yanagita-solvent}, and the generalized equipartition theorem holds for the such constrained systems \cite{kubo-book}.
Therefore, we adapt the equipartition theorem by simply replacing the active \HDs.
We derive the energy per \HD in \QE as
\begin{equation}
\qav{\fe}
=\frac{E_0}{\mf_{\rm sys}^\quasi}
=\frac{E_0}{\mf_{A}-2\mf_{Z}+\mf_{B}}.
\end{equation}

Then, the total energy attributed to the subsystem $A$ and $B$ are 
\begin{eqnarray}
\qav{\fE^{A}}
&=& \mf_{A}^\quasi\qav{\fe},\\
\qav{\fE^{B}}
&=& \mf_{B}^\quasi\qav{\fe},
\end{eqnarray}
respectively, where $\mf_{A}^\quasi=\mf_{A}-2\mf_{Z}$ and $\mf_{B}^\quasi=\mf_{B}$.
A schematic of the energy distribution in \QE is shown in the middle panel of Fig.~\ref{fig:schematic-equipartition}.
The total kinetic energy of the subsystem A becomes $\qav{E_{K_A}}=(\mf_{K_A}-\mf_{Z})\qav{\fe}$.
Because we observe kinetic energy per bead, we divide the total kinetic energy by the number of kinetic \HDs.
Therefore, the kinetic energy per bead and particle for subsystems $A$ and $B$ are
\begin{eqnarray}
\qav{\fe_{K_A} }
&=& \frac{\mf_{K_A}-\mf_{Z}}{\mf_{K_A}}\qav{\fe}\\
\qav{\fe_{K_B}}
&=&\frac{\mf_{K_B}^\quasi}{\mf_{K_B}} \qav{\fe}\nonumber\\
&=&\qav{\fe},
\end{eqnarray}
respectively.
Because the subsystem $B$ does not have frozen degrees, the number of \HDs in the kinetic parts in \QE is the same as that in equilibrium, i.e., $\mf_{K_B}=\mf_{K_B}^\quasi$, and $\mf_{Z}<\mf_{K_A}$,
\begin{eqnarray}
\qav{\fe_{K_A}} 
&=& (1-\frac{\mf_{Z}}{\mf_{K_A}})\qav{\fe}\\
&<&\qav{\fe} 
=\qav{\fe_{K_B}}
\end{eqnarray}
Therefore, the distributed energy per \HD for subsystem $A$ with a frozen degree is always smaller than that without the frozen degrees.
A schematic is shown in the bottom panel of Fig.~\ref{fig:schematic-equipartition}.
Further, we obtain the ratio as 
\begin{equation}
\frac{\qav{\fe_{K_A} }}{\qav{\fe_{K_B}}}
=1-\frac{\mf_{Z}}{\mf_{K_A}}.
\end{equation}
This means that the ratio does not depend on the number of \HDs for subsystem $B$.
For bead-spring systems, since $B$ corresponds to the solvent, we conclude that the kinetic energy of each solvent particle is always larger than that of the molecular beads in \QE.
We adopt these results for the chain and network molecules in the solvent, and we verify the numerical results.

\begin{figure*}[tb]
\begin{center}
\resizebox{\doublefiguresize}{!}{\includegraphics{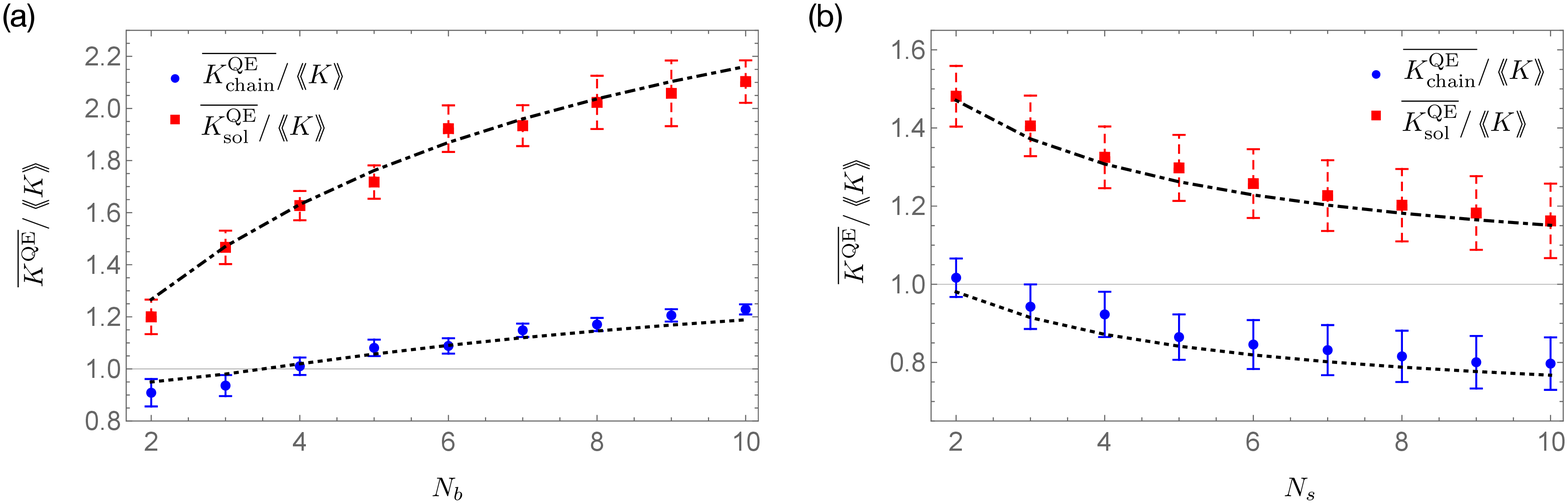}}
\end{center}
\vspace{-5mm}
\caption{
(a) Time-averaged kinetic energies for beads in chain molecules and solvent particles normalized by their equilibrium values as a function of $N_b$ are shown by circles and squares with error bars, respectively.
The dashed and the dotted-dashed lines are the theoretical estimations obtained using Eq.~(\ref{eq:theory-chain-kchain2}) and (\ref{eq:theory-chain-ksol2}) with $N_s=2$ and $\gamma=1/4$, respectively.
(b) Time-averaged kinetic energies for beads and solvent particles normalized by their equilibrium values are shown as a function of the number of solvent particles $N_{s}$. 
The dashed and the dotted-dashed lines are the same as in (a) with $N_b=3$.
$k=1000$, and $M=10$, and the other parameters are the same as those in Fig.~\ref{fig:chain_nsdep}.
}
\label{fig:chain_dep2}
\end{figure*}

\subsection{Averaged kinetic energy for chain molecule in quasi-equilibrium}
\label{subsec:equipartition-chain}

We apply the above theory to the chain molecule.
In the case of the chain molecule in the solvent, the system can be divided into a chain molecule corresponding to subsystem $A$ and the solvent particles corresponding the subsystem $B$.
The number of \HDs for chain molecule in the equilibrium is 
\begin{equation}
\mf_{\rm chain}=2N_{b}+(N_{b}-1),
\label{eq:HD-chain-eq}
\end{equation}
where $N_{b}$ and $N_{s}$ are the number of beads in the molecule and that of the solvent, respectively.
The first term corresponds to the number of \HDs for the kinetic part of the molecular beads, and the second term corresponds to the number of \HDs for the potential part of the molecule. 
The number of \HDs for the solvent is
\begin{equation}
\mf_{\rm sol}=2 N_{s}+\gamma N_{s},
\label{eq:HD-sol-eq}
\end{equation}
where the first and second terms correspond to the kinetic and potential parts of the solvent, respectively. 
Because the interaction between the solvent and molecule has a cut off distance $\ell_\alpha$, as in Eq.~(\ref{eq:potential-int}), it cannot be treated by the potential energy as a harmonic form.
Thus, we estimate the interaction energy as a linear function of the number of solvent particles with coefficient $\gamma$.
Therefore, using Eq.~(\ref{eq:HD-chain-eq}) and (\ref{eq:HD-sol-eq}), the energy per \HD in the equilibrium is given by
\begin{equation}
\eav{\fe}=\frac{E_0}{\mf_{\rm sys}}
=\frac{E_0}{3N_{b}-1+(2+\gamma)N_{s}},
\end{equation}
where $\mf_{\rm sys}=\mf_{\rm chain}+\mf_{\rm sol}$ is the total number of estimated \HDs for the system.
In \QE, the energy exchange between the vibration of the molecule and the other part is prevented.
The number of stiff springs is $\mf_Z=N_b-1$, and thus, the number of degrees of motion of the chain molecule is reduced to $\mf_{K_{\rm chain}}^\quasi=2N_b-(N_b-1)$ because the distance between the beads in a molecule is almost constant in \QE.
In other words, each spring acts as a rigid link, and the molecule behaves as if the motion has $N_b-1$ constraints.
Thus, the total number of frozen \HDs is $2\mf_{Z}=2(N_{b}-1)$ and $\mf_{\rm sys}^\quasi=\mf_{\rm chain}-2\mf_{Z}+\mf_{\rm sol}$.
The energy distributed to these \HDs is
\begin{equation}
\qav{\fe}
=\frac{E_0}{\mf_{\rm sys}^\quasi}=\frac{E_0}{N_{b}+1+(2+\gamma)N_{s}}.
\end{equation}
Therefore, the kinetic energies for chain beads and solvent particles can be estimated as
\begin{eqnarray}
\qav{\fe_{\rm chain}}
&=&\frac{E_0}{N_{b}+1+(2+\gamma)N_{s}}\Big(\frac{N_{b}+1}{3N_{b}-1}\Big) \label{eq:theory-chain-kchain1}\\
\qav{\fe_{\rm sol}}
&=&\frac{E_0}{N_{b}+1+(2+\gamma)N_{s}},\label{eq:theory-chain-ksol1}
\end{eqnarray}
and the quasi-equilibrium to equilibrium ratios for these kinetic energies are
\begin{eqnarray}
\frac{\qav{\fe_{\rm chain}}}{\eav {\fe}}
&=& \frac{N_{b}-1}{2N_{b}} \qav{\fe} \nonumber \\
&=& \Big(\frac{\qav{\fe_{\rm sol}}}{\eav {\fe}}\Big)\Big(\frac{1}{2}+\frac{1}{2 N_{b}}\Big) \label{eq:theory-chain-kchain2} \\
\frac{\qav{\fe_{\rm sol}}}{\eav {\fe}}
&=& 1+\frac{2(1-N_{b})}{N_{b}+1+(2+\gamma)N_{s}}.\label{eq:theory-chain-ksol2}
\end{eqnarray}
Comparisons between these analytic expressions and numerical experiments are shown in Fig.~\ref{fig:chain_dep2}.
In these figures, the theoretical expressions are indicated by the dashed and dotted-dashed lines.
The dashed lines correspond to the theoretical estimations using  Eq.~(\ref{eq:theory-chain-kchain2}) with $\gamma=1/4$, and the dotted-dashed lines correspond to the theoretical estimation using Eq.~(\ref{eq:theory-chain-ksol2}) with $\gamma=1/4$.

The time-averaged kinetic energy in the \QE dependency on the number of molecular beads $N_{b}$ and solvent particles $N_{s}$ is in good agreement with the theoretical estimations.
The ratio $\qav{\fe_{\rm sol} }/\qav{\fe_{\rm chain}}$ does not depend on the number of solvent particles and the parameter $\gamma$, that is,
\begin{equation}
\frac{\qav{\fe_{\rm sol} } }{\qav{\fe_{\rm chain}}}
=2-\frac{2}{N_{b}+1}.
\label{eq:theory-chain-ratio1}
\end{equation}
This means that the kinetic energy of the solvent particle is twice that of the long chain molecule for any number of solvent particles, and
for a long chain molecule with any fixed $N_{s}$, we get
\begin{eqnarray}
\lim_{N_{b} \rightarrow \infty} \frac{\qav{\fe_{\rm chain}}}{\eav{\fe}}
&=&\frac{3}{2}. \nonumber\\
\lim_{N_{b} \rightarrow \infty} \frac{\qav{\fe_{\rm sol}}}{\eav{\fe}} 
&=&3 \nonumber 
\end{eqnarray}
The kinetic energy per \HD for both the chain bead and solvent particles in \QE is always higher than that for the equilibrium values.
For a large number of solvent particles with fixed $N_{b}$, we derive that
\begin{eqnarray}
\lim_{N_{s} \rightarrow \infty} \frac{\qav{\fe_{\rm chain}}}{\eav{\fe}}
&=&\frac{2}{3}.\nonumber\\
\lim_{N_{s} \rightarrow \infty} \frac{\qav{\fe_{\rm sol}}}{\eav{\fe}}
&=&1 \nonumber
\end{eqnarray}
The kinetic energy of the solvent particles in \QE converges to the equilibrium value as $N_s$ increases, while the kinetic energy of the chain bead in \QE is always lower than that of the equilibrium value.

\begin{figure*}[tb]
\begin{center}
\resizebox{\doublefiguresize}{!}{\includegraphics{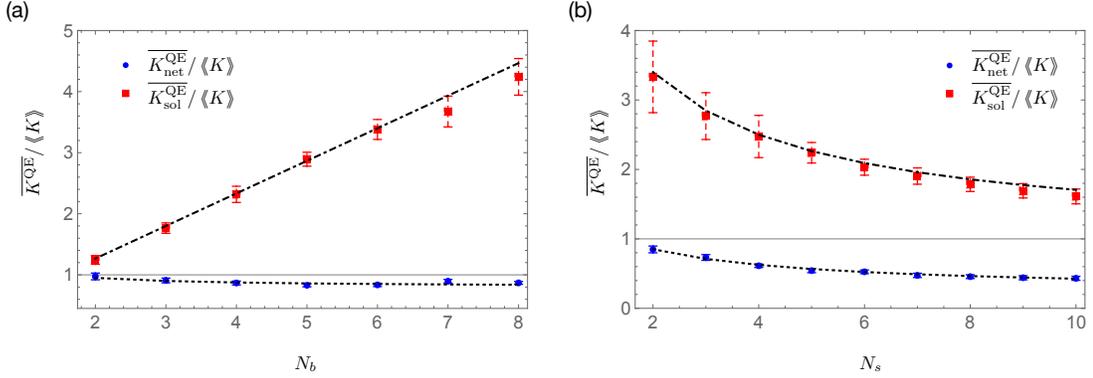}}
\end{center}
\vspace{-5mm}
\caption{
(a) Time-averaged kinetic energies for beads in network molecules and solvent particles normalized by their equilibrium values as the function of the number of beads $N_{b}$ are shown by circles and squares with error bars, respectively. 
The dashed and the dotted-dashed lines represent the theoretical estimations using Eqs.~(\ref{eq:theory-net-knet2}) and ~(\ref{eq:theory-net-ksol2}) with $N_{s}=2$.
(b) The same as in Figure (a) with the function of the number of solvent particles $N_{s}$. 
The dashed and dotted-dashed lines are the same as in figure (a) with $N_{b}=6$.
$E_0=0.1$.
The other parameters are the same as those in Fig.~\ref{fig:chain_nsdep}.
}
\label{fig:cgrf_dep2}
\end{figure*}
\subsection{Averaged kinetic energy for network molecule in quasi-equilibrium}
\label{subsec:equipartition-network}

We consider the theoretical estimation of the kinetic energy per \HD for the network molecule.
Here, we consider the complete graph as a connection topology for the molecule; the number of springs in the molecule is $_{N_{b}}C_{2}=N_{b}(N_{b}-1)/2$.
Thus, the number of springs is larger than that of the beads when $N_{b}$ is larger than three.
When these springs are stiff, the motion of the network molecule is restricted by the many springs, and it behaves as one particle.
Therefore, only the translational and rotational degrees of freedom remain in \QE; thus, $\mf_{\rm net}^\quasi=3$.
%
The number of independent \HDs for the potential energy of the springs in the molecule can be considered in the following manner.
Let us consider a simple scenario wherein the two beads are connected by two different springs with spring constant $k_1$ and $k_2$.
In this case, the two springs are not independent; rather, they behave as one spring with a spring constant is $k_1+k_2$.
Thus, the number of independent \HDs for such a molecule is five, to which the energy is equally distributed, i.e., four kinetic \HDs and one potential \HD for these springs.
In the case of a molecule which comprising four beads, and any pair of beads is connected by a stiff spring.
The connection topology is a complete graph, and there are six springs in the molecule.
The motion of these springs is no longer independent, and there are four independent springs in this case.
For the molecule with a complete-graph connection topology, there are $N_{b}(N_{b}-1)/2$ springs; however, the number of independent \HDs entered in such a potential energy is
\begin{eqnarray}
\mf_{\rm spring}
&=&2N_{b}-3.
\label{eq:eff_spring}
\end{eqnarray}
The number of independent springs is the same as the degree of the framework, which is the rank of the rigid matrix in the literature \cite{Maxwell1864,Laman1970,Asimow1978,rigidity-book}.
In Appendix~\ref{subsec:mcmc}, we check the number of independent \HDs for network molecules via the direct calculation of the thermal average of the potential energy using the Markov chain Monte Carlo method. 

This estimation of the number of independent \HDs can be explained intuitively as follows.
When the spring constant is sufficiently large, we observe that springs act as constraints for the motion of the molecular beads, and the molecule behaves as one particle, as explained previously.
In fact, during \QE, the energy exchange between the springs and other parts is prevented, and thus, the potential energy for springs is almost zero which is the initial value.
Then, the total number of degrees for the network molecule is the transnational and rotational degrees.
Thus, the following equation holds, 
\begin{equation}
\mf_{\rm net}^{\quasi}
=2N_{b} -\mf_{Z}=3.
\end{equation}
where $\mf_{Z}=\mf_{\rm spring}$.
The number of \HDs of the solvent part is the same as in the case of the chain molecule, as explained in the previous subsection, 
$$
\mf_{\rm sol}=2 N_{s}+\gamma N_{s}.
$$
Therefore, the number of \HDs for the whole system can be estimated using
\begin{eqnarray}
\mf_{\rm sys}&=&2N_{b}+\mf_{\rm spring}+\mf_{\rm sol} \nonumber \\
&=&4N_{b}-3+(2+\gamma)N_{s}\\
\label{eq:net_freedom1}
\mf_{\rm sys}^\quasi&=&2N_{b}-\mf_{Z}+\mf_{\rm sol}\nonumber\\
&=&3+(2+\gamma)N_{s}.
\label{eq:net_freedom2}
\end{eqnarray}
Thus, the distributed energy per \HD during \QE given as
\begin{equation}
\qav{\fe}=\frac{E_0}{\mf_{\rm sys}^\quasi}=\frac{E_0}{3+(2+\gamma)N_{s}}.
\end{equation}
Using the above estimation of the number of \HDs and assuming that the total energy is distributed equally to these degrees, we obtain that the average energy per \HD as
\begin{eqnarray}
\qav{\fe_{\rm net}}&=&\qav{\fe}\frac{\mf^\quasi_{\rm net}}{\mf_{\rm net}} \nonumber\\
&=&  \frac{3 E_0}{\{3+(2+\gamma)N_{s}\}(4N_{b}-3)}
\label{eq:theory-net-knet1}\\
\qav{\fe_{\rm sol}}&=&\qav{\fe} \nonumber \\
&=& \frac{E_0}{3+(2+\gamma)N_{s}}
\label{eq:theory-net-ksol1}
\end{eqnarray}
and the ratios to its equilibrium value $\eav{\fe}=E_0/\mf_{\rm sys}$ are
\begin{eqnarray}
\frac{\qav{\fe_{\rm net}}}{\eav{\fe}}
&=&\left\{1+\frac{4N_{b}-6}{3+(2+\gamma)N_{s}}\right\} \frac{3}{2N_{b}} 
\label{eq:theory-net-knet2} \\
\frac{\qav{\fe_{\rm sol}}}{\eav{\fe}}
&=&\left\{1+\frac{4N_{b}-6}{3+(2+\gamma)N_{s}}\right\} 
\label{eq:theory-net-ksol2}
\end{eqnarray}
The time-averaged energy dependency on the number of beads $N_{b}$ and solvent particles $N_{s}$ are plotted with these theoretical estimations, as shown in Figs.~\ref{fig:cgrf_dep2} (a) and (b), respectively. 
The dashed and dotted-dashed lines in these figures represent the theoretical estimations by Eq.~(\ref{eq:theory-net-knet2}) and (\ref{eq:theory-net-ksol2}) with $\gamma=1/4$, respectively.
These theoretical estimations are in good agreement with the numerical results.

In the case of the network molecule, the ratio of these averaged energies is inversely proportional to $N_{b}$ as 
\begin{equation}
    \frac{\qav{\fe_{\rm net}}}{\qav{\fe_{\rm sol}}}=\frac{3}{2N_{b}},
    \label{eq:theory-net-ratio1}
\end{equation}
and does not depend on the parameter $\gamma$.
This implies that for a molecule composed of a large number of atoms (e.g. a biomolecule), the averaged kinetic energy of such a molecule becomes considerably lower than that of a solvent particle during \QE.

For a large number of solvent particles with a fixed number of beads, we get
$$
\lim_{N_{s}\rightarrow\infty} \frac{\qav{\fe_{\rm net}}}{\eav{\fe}}=\frac{3}{2N_{b}}.
$$

$$
\lim_{N_{s}\rightarrow\infty} \frac{\qav{\fe_{\rm sol}}}{\eav{\fe}}=1,
$$

These ratios also do not depend on the parameter $\gamma$.

\section{Summary}
~\label{sec:summary}
We investigated the emergence of the quasi-equilibrium (\QE) over the course of  relaxation to equipartition for Hamiltonian dynamics with fast and slow time scales.
We used the bead-spring model as a simple model of molecules and analyzed the energy distribution of molecules and solvents during the \QE state.
The model molecule comprises of the masses (beads) connected by linear springs in a two-dimensional space, which interact with solvent particles.
Further, we considered the initial condition that a molecule has only the translational kinetic energy without fast vibration, which means that the potential energy of the springs between the beads in the molecule is initially set to zero.
For such initial conditions, the numerical simulations showed that all kinetic energies for molecular beads and solvent particles become equal on average, meaning that they relax to equipartition.
However, the relaxation time to equipartition depends on the stiffness of the spring between the beads, which determines the time scale of the molecular vibration.
The relaxation time can be extremely long when the time scale of the molecular vibration is shorter than that of the collision between the molecule and solvent particles.
The relaxation time obeys the Boltzmann--Jeans law \cite{konishi-yanagita-solvent}.
During such a long transition to equipartition, we observed that the time-averaged kinetic energy for molecular beads and solvent particles was almost constant for a long time.
We call this state the quasi-equilibrium (\QE).

During the \QE, we observed that the time-averaged kinetic energy of a solvent particle is always larger than that of the bead irrespective of the type of molecular structure (we considered two types of molecules: chain and network).
In other words, the solvent particles behave more energetically than the molecular beads in the \QE.
We numerically show the time-averaged kinetic energy dependency on the number of molecular beads and solvent particles.
For chain molecules, both the molecular and solvent kinetic energies decrease as the length of the chain molecule increases.
However, a network molecule, the solvent kinetic energy takes a constant value as the number of beads in the molecule increases.
The ratio of molecule energy to solvent energy increases with an increase in the number of beads for both types of molecules.
We state that the ratio does not depend on the number of solvent particles, which is usually large in reality.
The functional forms of these dependencies are different, and the dependency is determined by the connection topology of the molecule.
To clarify these dependencies, we adapt the equipartition theorem for \QE by considering ``frozen degrees.''

In equilibrium, the equipartition theorem states that the thermal average of the energy for all ``harmonic'' degrees (\HDs) is equal.
Here, the number of the \HDs denotes the number of independent harmonic terms that enter the Hamiltonian of the system. 
We explained that the numerical results observed in \QE can be understood by adapting the equipartition theory and by estimating the effective number of \HDs. 
The theoretical analysis is in good agreement with the numerical results.

In the case of a chain molecule, the molecular vibration is frozen, and the motion of the molecule behaves as if the distance between the beads is constant. 
Then, the number of frozen \HDs is $2(N_b-1)$ where $N_b$ denotes the number of beads in the chain molecule.
We showed that the kinetic energy of the chain molecule in \QE is lower than that in equilibrium for the short-length molecule $N_b \leq 3$; further it becomes higher when $N_b >3$.
For the long chain, the energy of solvent is double that of the molecule in \QE.
The solvent's kinetic energy in \QE is triple that in equilibrium.
These results show that that solvent behaves in a more energetic manner than molecule with frozen degrees.

In the case of network molecules, we determined that the energy ratio increases linearly with the number of beads in the molecule by estimating the number of ``independent'' springs in the molecule.
This implies that large molecules with complex connection topology, such as bio-molecules, are less energetic than solvents.
When the number of solvent particles is larger, the kinetic energy per \HD of the molecule in \QE equals that of the equilibrium value.
The energy of the molecule depends linearly on the inverse of the number of beads in the molecule.
Both facts suggest that the energy ratio in \QE does not depend on the number of solvent particles, which is usually large in reality.
This suggests that \QE can be observed by measuring the kinetic energy ratio, and it indicates the value of ``frozen degrees'' that exist in the system in \QE.

The proper functioning of biomolecules occurs out of the equilibrium.
Recent NMR studies showed that the high mobility of the terminal residues appear to change only by an amino acid \cite{Miura2020}.
The change in the amino acid causes a helical conformation, and the hydrogen bonds are regarded as ``frozen degrees.''
Thus, introducing the ``frozen degrees'' may induce differences in the mobility.
Because \QE lasts a long time, we can observe the difference in the average kinetic energy in the experiments, and through the kinetic energy ratio of the molecule to solvent, we can estimate how many frozen or resting degrees of freedom exist in the molecule.
Such an inhomogeneity of energy distribution emerging from the frozen degree may be an indicator of how far it is from equilibrium, and it can shed some light on the functioning of the biomolecule.

\acknowledgments
The authors would like to thank Professor M. Toda and Dr. Y. Y. Yamaguchi for the fruitful discussions.
This work was supported by JSPS KAKENHI Grant Numbers 24540417 and 15K05221. 
One of the authors (T.K.) would like to thank Chubu University for the financial support. 

\appendix
\section{Estimation of the harmonic degrees using the Markov Chain Monte Carlo}
\label{subsec:mcmc}
In Subsec.~\ref{subsec:equipartition-chain} and ~\ref{subsec:equipartition-network}, we adapt the equipartition theorem to estimate the energy distribution during the \QE state.
We use the harmonic degrees (\HDs) in which the energy is distributed equally to such degrees in \QE.
In the case of a chain molecule, the number of frozen motions originates from that of the springs in the molecule.
However, for the bead-spring network molecule, if the number of springs exceeds $2N_{b}-3$ the polygon can be triangulated.
This means that we cannot change the spring length independently, i.e., a change in the length of the spring leads to a change(s) in the other length(s).
Thus, the potential energy of all springs is not a simple sum of the harmonic terms of the springs for such network molecules.
In Subsec.~\ref{subsec:equipartition-network}, we estimated the number of independent \HDs as $\mf_{\rm spring}=2N_{b}-3$ for network molecules with the complete-graph connection topology.
This is the same as the degree of a framework, which is the rank of a rigid matrix in the literature \cite{Maxwell1864,Laman1970,Asimow1978,rigidity-book}.

\begin{figure}[t]
\vspace{2mm}
\begin{center}
\resizebox{\singlefiguresize}{!}{\includegraphics{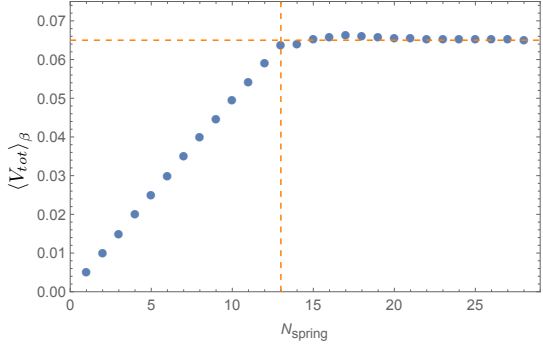}}
\end{center}
\vspace{-5mm}
\caption{Thermal-average of the total potential energy of springs for randomly connected network molecules as a function of the number of springs.
The thermal average is calculated by MCMC.
The vertical dashed line represents the triangulation threshold above which the total potential energy
is not a sum of independent \HDs.
The horizontal dashed line indicates the maximum total potential energy, i.e., $\frac{2N_{b}-3}{2 \beta}$.
The parameters are $N_{b}=8$, $\beta=1$, and $M=500$.
}
\label{fig:mcmc1}
\end{figure}

In this appendix, we verify the number of the independent \HDs for randomly connected network-molecules using Markov chain Monte Carlo (MCMC) in equilibrium.
The randomly connected network-molecules with a fixed number of springs are simply generated in the following manner.
The beads are located at an angle of the equilateral polygon placed on the unit circle, and the randomly selected $N_{\rm spring}$ pairs of beads are connected by a spring; duplicate pairs are forbidden.
That is, the spring coefficient is $k_{i,j}=k$ in Eq.~(\ref{eq:hamiltonian-net2}) if $i$ and $j$ beads are connected; otherwise, $k_{i,j}=0$.

The thermal average of the total spring energy $\cav{ V_{tot}}$ for a randomly connected network molecule with a fixed $N_{\rm spring}$ is
\begin{equation}
\cav{V_{tot}}
=\cav{ \sum_{i=1}^{N_{b}-1} \sum_{j = i+1}^{N_{b}}\frac{k_{i,j}}{2}\left(\left| \VEC{r_{j}}-\VEC{r_i}\right| -\ell_{i,j}\right)^2 }
\end{equation}
calculated by MCMC \cite{Newman99,Landau05}.
Further, we average the total spring energy over $M$ different random networks.
Fig.~\ref{fig:mcmc1} shows the thermal average of the total potential energy as a function of the number of springs with $N_{b}=8$ is shown.
The thermal average $\cav{V_{tot}}$ increases linearly with the number of springs, whereas the number is below the regularization threshold.
Above the threshold, the total potential energy takes a constant value, i.e., $\frac{2N_{b}-3}{2 \beta}$.
Therefore, the number of independent \HDs for which the energy is equally distributed does not increase as the number of springs increases.
The MCMC result is consistent with the previous estimation that the number of independent \HDs is $2N_{b}-3$ for the complete graph.

\bibliography{refchain}

\end{document}